\documentclass[preprint,12pt]{elsarticle}

\usepackage[margin=2.cm]{geometry}

\usepackage{epsfig} 
\usepackage{graphicx} 

\usepackage[rightcaption]{sidecap}

\usepackage{hyperref}   
\hypersetup{
	colorlinks=true,
	linkcolor=blue,
	citecolor=blue,
	urlcolor=blue,
        pdfauthor=author
}

\usepackage{amssymb}
\usepackage{amsmath}
 \usepackage{array}

 \newcolumntype{M}[1]{>{\centering\arraybackslash}m{#1}}

\usepackage{setspace}

\journal{Mechanical Systems and Signal Processing}
\begin{document}

\begin{frontmatter}

\title{Pivot Bearings for Efficient Torsional Magneto-Mechanical Resonators} 

\author[1]{Chengzhang Li}
\author[1]{Ali Kanj}
\author[1]{Jiheng Jing}
\author[1]{Gaurav Bahl}
\author[1,2]{Sameh Tawfick\corref{cor5}}
\cortext[cor5]{Corresponding author. Email: tawfick@illinois.edu}

\affiliation[1]{organization={Department of Mechanical Science and Engineering},
            addressline={University of Illinois at Urbana-Champaign}, 
            city={Urbana},
            state={IL},
            postcode={61801},
            country = {USA}}
            
\affiliation[2]{organization={The Beckman Institute for Advanced Science and Technology},
            addressline={University of Illinois at Urbana-Champaign}, 
            city={Urbana},
            state={IL},
            postcode={61801},
            country = {USA}}

\begin{abstract} 
{
Rotating magnets have recently emerged as an efficient method for producing ultra-low frequency signals for through-earth and through-seawater communications. Magneto-mechanical resonator (MMR) arrays, which are magnetized torsional rotors with a restoring torque, are a promising implementation of this idea that use resonance to enhance the magnetic signal generation.
The fundamental challenge for MMR design is to have a suspension system for the rotors capable of resisting large transverse magnetic forces while allowing for a large angle of motion at low dissipation and hence high efficiency.
Here, we study flexure-based pivot bearings as compliant support elements for MMR rotors which address this challenge and demonstrate their efficient low damping operation. A crossed-flexure configuration enables large angular rotation around a central axis, large transverse stiffness, and compact assembly of closely spaced rotor arrays via geometric flexure interlocking. We characterize the eigen frequency performance and structural damping of MMR supported by these proposed pivot bearings. We develop analytical and numerical models to study their static and dynamic behaviors, including their coupled dynamic modes. We demonstrate that their damping coefficient is up to 80 times lower than corresponding ball bearing MMR. This study is broadly applicable to various systems that leverage arrays of coupled torsional oscillators such as magneto-mechanical transmitters, metamaterials, and energy harvesters.
}
\end{abstract}  


\end{frontmatter}
\begin{singlespace}
\section{Introduction}

Mechanical resonators with torsional degrees of freedom are found in a variety of engineering applications including metamaterials~\cite{Attarzadeh2019, Beli2018}, explorations of topology~\cite{Liu2022, Grinberg2020}, and magneto-mechanical transmitters~\cite{Thanalakshme2022, Kanj2022, Jing2023, Srinivas2019, Fereidoony2022}. As of 2024, there are more than 200 published studies on various magneto-mechanical resonator devices, as per a web of knowledge search of magneto-mechanical resonators using various keywords. In this article we focus our attention on magneto-mechanical resonator (MMR) arrays, which are composed of torsional mechanical rotors that possess a permanent magnetic dipole (PM), have a torsional resonance, and can couple together magnetically~\cite{Grinberg2019}. When undergoing periodic oscillations, the synchronized motions of MMR arrays have been discovered as an efficient approach to produce ultra-low frequency (ULF band) magnetic signals. This approach to producing ULF signals opens a communications window through high-conductivity media such as water, metal, soil, and rock, all of which are generally impenetrable to electromagnetic signals in radio frequency bands \cite{Akyildiz2005, Rowe1974, Wang2021, Dong2020, Burch2018, Barani2019}. 

The primary energy loss mechanism in an MMR-based transmitter, and the mechanism that determines its power efficiency, is the mechanical damping of the rotor motion that occurs in the suspension system \cite{Kanj2022, Jing2023, Srinivas2019, Madanayake2017}. Another major consideration for producing a large ULF signal with an MMR is the amplitude of the torsional oscillation that the suspension is capable of supporting. In addition, the suspension must resist the large linear forces occurring in the transverse direction between adjacent permanent magnetic rotors. In this context, suspension systems that maximize the transverse stiffness to angular stiffness ratio are extremely important. 

Ball bearings are commonly used in MMRs as suspension systems~\cite{Thanalakshme2022, Srinivas2019, Madanayake2017} because they readily satisfy these requirements. However in practice, they do experience significant mechanical dissipation~\cite{Thanalakshme2022, Jing2023} due to rolling friction, resulting in low power efficiency as will be detailed later in this manuscript. The dissipation due to rolling friction is possibly exacerbated due to the oscillating angular displacements, but this question remains to be investigated. Presently, however, the use of other types of low-loss frictionless bearings is of interest to improving the power efficiency of MMR-based devices including MMR transmitters. 

Flexible joints are mechanical elements potentially suitable to construct frictionless bearings, in which the flexible elements are arranged to allow the angular displacement while constraining the transverse displacement. Flexible joints can be divided into two varieties: notch-type springs and leaf springs~\cite{Trease2005, Smith2000}. Leaf springs have been used in a variety of ways to create revolute joints, such as the cartwheel hinge~\cite{Trease2005} and cross-strip hinge~\cite{Hopkins2010_FACTP1, Hopkins2010_FACTP2}. Compared with the cartwheel hinge at the same size, the cross-strip hinge provides higher transverse stiffness and more compliance in rotational direction (before yielding)~\cite{Trease2005}.

Commercial cross-strip bearings, sometimes known as pivot bearings, consist of flexural and rigid components meticulously assembled and joined through brazing or welding techniques~\cite{Flex-Pivots2022}. This intricate manufacturing process necessitates advanced equipment and specialized fabrication for each unique bearing design, resulting in a limited selection of compatible bearing materials and dimensions~\cite{C-Flex2019}. Significant challenges constrain the design space of existing cross-flexure configuration, geometries and materials. More specifically, to increase the maximum rotational angle, ultrathin or ultralong leaves are in bearing design and production procedure to tailor a cross-strip bearing according to our MMR's requirements poses significant challenges and comes with substantial costs. At the same time, the cylindrical configuration of the prevailing bearing design places a limitation on the space between adjacent rotors since the suspensions must not obstruct each other.

In this study, we design, fabricate, and test a flexural cross-strip hinge pivot bearing, customized for MMRs. We derive analytical equations for its mechanical compliance. We study its dynamic characteristics when used in an oscillating MMR. Eigenmodes, eigenfrequencies, and mode-crossing phenomena are described and analyzed in this dynamic analysis. We present pivot bearings made from various materials and dimensions and evaluate their use in MMR. The article is organized as follows: Section \ref{sec:Design} states the working principle of the MMR system, the design, and the constraints for the pivot bearing. Section \ref{sec:Analytical} describes the analytical model of each part in our MMR system, including both static and dynamic analysis. Section \ref{sec:Experimental} depicts the experimental setups. Section \ref{sec:Results} illustrates and analyzes the experimental results. The conclusions are presented in Sec. \ref{sec:Conclusion}.

\section{Design of Magneto-Mechanical Resonators}\label{sec:Design}

\subsection{MMR Working Principle}

\begin{figure}[htb]
\centerline{\psfig{figure=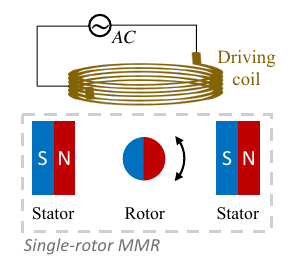,width=0.35\textwidth}}
\caption{Working principle of a single-rotor magneto-mechanical resonator (MMR), electromechanically actuated by a coil. A restoring torque results from the angular displacement of the rotor due to the arrangement of the stator magnets.}
\label{1rotorschematic.ps}
\end{figure}

A single-rotor magneto-mechanical transmitter system depicted in Fig. \ref{1rotorschematic.ps} comprises two main components: the single-rotor MMR and a driving coil~\cite{Thanalakshme2022, Kanj2022, Jing2023}. The MMR consists of a torsional resonator made from a cylindrical permanent magnet, which we call the rotor, and static permanent magnetic plates, referred to as the stators. The cylindrical magnetic rotor is held at the center of the frame by two pivot bearings, allowing for only one rotational degree of freedom (DoF). The rotor is magnetized perpendicular to the longitudinal rotation axis of the cylinder (transverse magnetization), and the stator is magnetized along its shortest thickness direction. At equilibrium, the dipole moments of the rotor and stators have the same orientation. The stators provide an external magnetic flux that induces a restoring torque when the rotor deviates from its equilibrium position. Additional restoring torque is provided by the flexure elastic deflection in the pivot bearings (see Fig. \ref{PivotBearings.ps}). A coil placed next to the frame drives the torsional oscillator by generating an alternating magnetic flux to perturb the rotor from its initial stationary equilibrium. 

A multi-rotor magneto-mechanical transmitter system (see Fig. \ref{2rotorSchematic.ps}(a)) having two or more rotors can be constructed following the same working principle~\cite{Thanalakshme2022}. As the magnetic field generated by MMR is directly proportional to the total rotor volume, a multi-rotor configuration with a smaller moment of inertia (MoI) per rotor exhibits a higher resonance frequency compared to a single-rotor configuration with an equivalent total rotor volume, while maintaining the same magnetic field generation~\cite{Thanalakshme2022}. Additionally, since two magnetic stators need to be allocated for each device, a compact multi-rotor device with multiple rotors oscillating simultaneously in an array provides improved space efficiency.

\subsection{Pivot Bearing Working Principle}
The schematic of the new pivot bearing design is shown in Fig. \ref{PivotBearings.ps}. Like any other type of flexure hinge, the rotational DoF is given by the deflection of the flexures, without any frictional or rolling contact between moving parts. As discussed later, this new design tackles a known drawback of pivot bearings in general, which is the limited angular rotation necessary to prevent material yielding in the flexure~\cite{Wu2015}. Fig. \ref{PivotBearings.ps}(a) shows the geometric parameters of a single flexure layer. The thin flexure of length $l$ plays a vital role in ensuring the operational functionality of the pivot bearing. We consider the bearing layer as a thin cantilever beam to derive insights and mathematical modeling. With one side of the cantilever anchored and an impulse excitation acting on the free end along the thickness direction, in accordance with the first mode of vibration, the free end of the cantilever beam oscillates periodically around a center point~\cite{Smith2000}. We introduce an end cap to connect the free end of the cantilever beam with the rotor in order for the rotating center of the beam to align with the pivot point of the rotor (see Fig. \ref{PivotBearings.ps}(c)). In such a way, the resonator rotates only around the designated oscillating axis without deviating from its pivot. Additional beams, sharing the same rotating center, could be installed to increase the lateral stability by maximizing the stiffness in the transverse direction (see Fig. \ref{PivotBearings.ps}(d)) of the mechanism.

\subsection{Pivot Bearing Design Parameters}\label{sec:design_constraints}

\begin{figure*}[htb]
\centerline{\psfig{figure=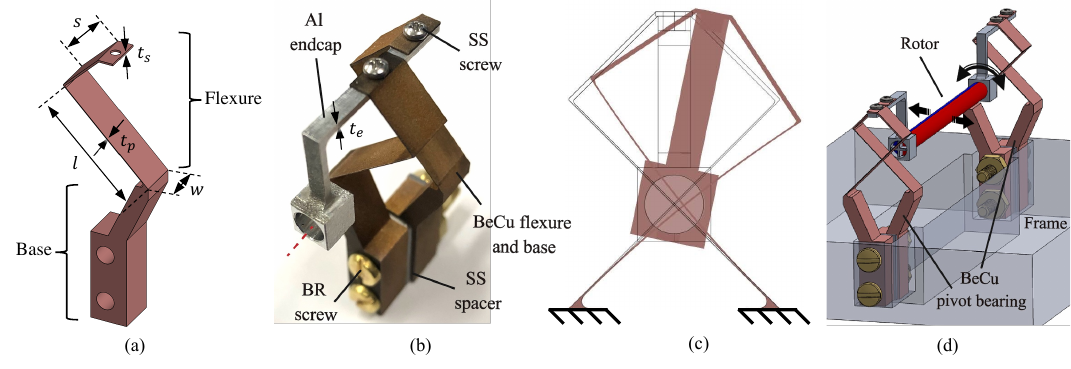,width=0.95 \textwidth}}
\caption{Pivot bearing design and construction. (a) Isometric view of the pivot bearing flexure layer. The 'Flexure' section refers to the compliant beam, while the 'Base' section remains rigid and non-deformable. (b) Isometric view of the copper beryllium (BeCu) Pivot Bearing: comprised of two BeCu base and flexure layers, one aluminum (Al) end cap, one stainless steel (SS) spacer, two SS screws, and two brass (BR) screws. The red dashed line indicates the rotational axis of the rotor (not shown) and the bearing. (c) Orthogonal view along the pivot bearing rotation axis depicting the angular vibration mode shape, with the flexure ends fixed at the bottom (the 'Base' section is omitted). The black edges represent the undeformed condition of the bearing. The center of the cylindrical slot in the endcap remains aligned with the rotational pivot throughout the oscillation. (d) Schematic of BeCu pivot bearing rotor assembly: one cylindrical PM connected by two BeCu pivot bearings, fastened to a PLA frame (gray box) using BR screws. The transverse arrows show the forces on the rotor from the stators (omitted). The circular arrows show the restoring torque.}
\label{PivotBearings.ps}
\end{figure*}

During the design process, it is essential to take into account various constraints, including manufacturing limitations, material selection, and structural stability requirements. The bearing layer is manufactured using a Wire-EDM machine, allowing for the production of thin beams with a minimum thickness of 0.1~mm and a maximum length-to-thickness aspect ratio of 200. Hence, we choose a flexure thickness $t_p$  of 0.1~mm for the pivot bearing shown in Fig. \ref{PivotBearings.ps}(a). The flexure length $l$ is adjusted between 10~mm and 20~mm based on the specific requirements of the MMR.

The primary sources of pivot bearing dissipation are the flexure material intrinsic loss and anchor losses~\cite{Bindel2005}. Specifically, materials with lower intrinsic damping dissipate less energy per cycle, resulting in a higher quality factor (Q-factor). Meanwhile, materials with high ductility and yield strength are appropriate to improve the rotor's range of rotation. Considering manufacturability and inherent material damping~\cite{ASHBY2006}, we selected stainless steel (SS), aluminum (Al), brass (BR), and copper beryllium (BeCu) as candidate materials for study. BeCu is known to exhibit low dissipation and hence was an attractive candidate for this application.

\begin{figure}[htb]
\centerline{\psfig{figure=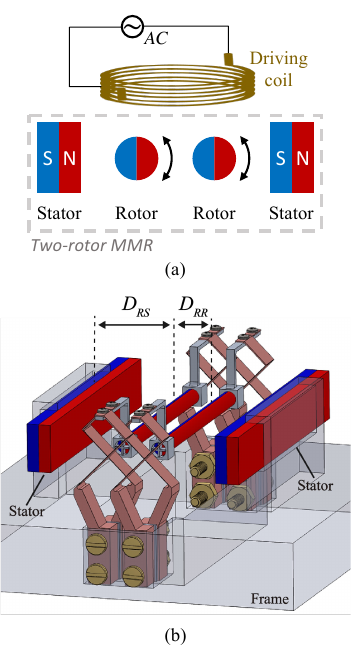,width=0.4\textwidth}}
\caption{(a) Schematic of two-rotor MMR, with the coil electromechanical actuation. (b) ISO view of Pivot Bearing two-rotor MMR with an interlocked configuration of adjacent pivot bearings. Pivot bearings and PM stators are secured onto a PLA frame. The schematic also displays the rotor-to-rotor (RR) distance, denoted as $D_{RR}$, and the rotor-to-stator (RS) distance, denoted as $D_{RS}$.}
\label{2rotorSchematic.ps}
\end{figure}

During MMR device operation, the PM rotors are subject to large magnetic forces from the coil and stators. The system can become unstable in the lateral (transverse) direction (as shown in Fig. \ref{PivotBearings.ps}(d)) if the reaction forces are inadequate to hold the rotors in place. Hence, the pivot bearings need to have large lateral stiffness to resist the large magnetic forces. In the proposed design, two mirror-symmetric flexure layers of thickness $w$ are assembled within each bearing to increase the lateral stiffness. Additionally, the resonance frequency of a multi-rotor MMR is controlled by the magnetic rotor-to-rotor (RR) distance~\cite{Thanalakshme2022}, which is described in the next section. The pivot bearing layer is given a zigzag shape to enable better lateral packing as shown in Fig. \ref{2rotorSchematic.ps}(b), resulting in shorter RR distances and ensuring that the synchronously oscillating rotors do not interfere with each other. 

Considering the constraints above, the flexure layer is designed and manufactured as the schematic shown in Fig. \ref{PivotBearings.ps}(a). The final shape of the pivot bearing assembly (see Fig. \ref{PivotBearings.ps}(b)) consists of three components: two mirrored stacking flexure layers, the L-shaped end cap, and the fixing accessories (screws and spacers). The aluminum L-shaped end cap aligns the rotor center with the bearing's rotating axis, allowing nearly pure pivoting around the rotor's center axis. Two AISI 0-80 screws fasten the end cap to the two flexure layers, while a SS spacer of 0.04~in (1.016~mm) thick separates two flexure layers, preventing collisions during bearing oscillation. Fig. \ref{PivotBearings.ps}(d) shows a complete pivot bearing rotor assembly. The cylindrical magnet is affixed to the end caps using adhesive, while the ends of the pivot bearings are fastened to the 3D printed plastic frame (made of polylactic acid (PLA)) using BR 4-40 screws and bolts. 

\begin{figure}[htb]
\centerline{\psfig{figure=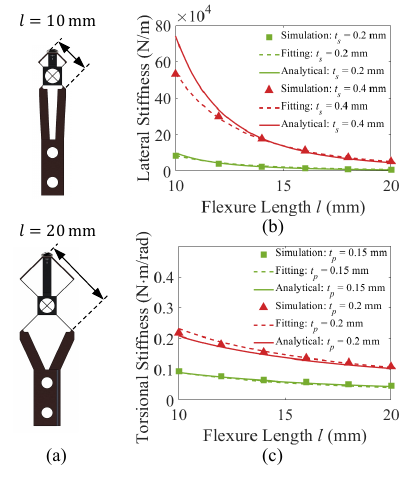,width=0.5\textwidth}}
\caption{(a) Schematic comparing the shape of $10$~mm length (top) and $20$~mm length pivot bearing (bottom). (b) BeCu Pivot bearing rotor assembly ($t_p = 0.1$~mm, $w = 6.35$~mm) lateral stiffness $K_{b}^{l}$ for $t_s = 0.2$~mm (green) and $t_s = 0.4$~mm (red): FEA simulation (marker), power-law fitting (dashed line), and analytical (solid line) with different flexure lengths $l$. (c) Pivot bearing rotor assembly ($t_s = 0.4$~mm, $w = 6.35$~mm) torsional stiffness for $t_p = 0.15$~mm (green) and $t_p = 0.2$~mm (red): FEA simulation (marker), power-law fitting (dashed line), and analytical (solid line) with different flexure lengths.}
\label{2sizeLSPB.ps}
\end{figure}

\section{Analytical Model}\label{sec:Analytical}
\subsection{Static Analysis: Magnetic Restoring Stiffness} \label{sec:rotorstatic}

First, we describe the restoring forces, torques, and stiffnesses on the rotor due to the presence of the stators. The magnetic dipole approximation is convenient for computing the magnetic field at the rotor's position generated from two symmetrically placed PM stators 
 
\begin{equation}\label{eq:magnetic_field}
B_{S} = \frac {\mu_0 m_{S}}{\pi {D_{RS}}^3},
\end{equation}
where $\mu_0$ is the vacuum permeability constant, $D_{RS}$ is the rotor-to-stator (RS) distance. $m_R$, $m_S$ are the magnetic moments of the rotor and stator, respectively, which can be calculated using the formula
\begin{equation}\label{eq:magnetic_moment}
m_{R,S} = \frac { B_r V_{R,S}}{ \mu_0 },
\end{equation}
where $B_r$ is the remanence. $V$ represents the volume of the PM, with subscripts $R$ and $S$ denoting the rotor and stator, respectively. 

The magnetic torque $T_{m}$ and lateral force $F_{m}$ between one rotor and two symmetrically placed stators can then be derived assuming a magnetic dipole approximation

\begin{equation}\label{eq:magnetic_load}
    \begin{cases}
    T_{m} = - \frac {\mu_0 m_{S}m_R}{ 2 \pi }\left((D_{RS} - x)^{-3} + (D_{RS} + x)^{-3}\right)\sin{\theta} \\[8pt]
    F_{m} = \frac {3\mu_0 m_{S}m_R}{ 2 \pi }\left((D_{RS}-x)^{-4}-(D_{RS}+x)^{-4}\right)\cos{\theta}
    \end{cases},
\end{equation}
where $\theta$ is the angular displacement of the rotor from its static equilibrium orientation, and $x$ represents the lateral displacement of the rotor from its static equilibrium position. 

Expanding Eq. (\ref{eq:magnetic_load}) around $\theta=0$ and $x=0$ using Taylor series and substituting Eq. (\ref{eq:magnetic_field}) yields
\begin{equation}\label{eq:magnetic_load_taylor}
    \begin{cases}
    T_{m} = - m_RB_{S}\left(\theta - \frac{1}{6}\theta^3 + \frac{6}{{D_{RS}}^2}x^2\theta + O\left((x,\theta)^4\right) \right) \\[8pt]
    F_{m} = \beta{m_S}^{-\frac{2}{3}}{m_R}{B_S}^\frac{5}{3}\left(x + \frac{5}{{D_{RS}}^2}x^3 -\frac{1}{2}x\theta^2 + O\left((x,\theta)^4\right) \right)
    \end{cases},
\end{equation}
where $\beta = 6\left(\frac{2^{5} \pi^2}{{\mu_0}^2}\right)^\frac{1}{3}$ is a constant term.
For small displacements, we can keep only the first-order term and disregard the high-order terms in Eq. (\ref{eq:magnetic_load_taylor}), thus deriving linear magnetic stiffnesses with respect to only rotational or translational displacement, which can be expressed as 

\begin{equation}\label{eq:magnetic_stiffness}
   \begin{cases}
K_{m}^{\theta} = m_RB_{S} \approx - \frac{T_{m}}{\theta} \\[8pt]
K_{m}^{l}=- \beta{m_S}^{-\frac{2}{3}}{m_R}{B_S}^\frac{5}{3} \approx - \frac{F_{m}}{x}
    \end{cases}.
\end{equation}

We note here that when the RS distance $D_{RS}$ is small, significant errors may arise from the magnetic dipole assumption without considering the geometry shape of the PMs~\cite{Jing2023}. Consequently, we use finite-element analysis (FEA) to calculate the magnetic field experienced by each PM~\cite{Salon1995} to more accurately determine $T_{m}$ and $F_{m}$. A comparison of the magnetic field calculated using the dipole approximation Eq. (\ref{eq:magnetic_moment}) and FEA as a function of $D_{RS}$ is demonstrated in [supplementary material Sec. S6]. Accordingly, the stiffness $K_{m}^{l}$ and $K_{m}^{\theta}$ shown in Eq. (\ref{eq:magnetic_stiffness}) can also be numerically estimated from the change of load by a small prescribed perturbation in position and angle using FEA, respectively. 

\subsection{Static Analysis: Pivot Bearing Restoring Stiffness}\label{sec:bearingstatic}
\subsubsection{Analytical Equation}

When the flexure length $l$ is much greater than its thickness $t_p$, the shape coefficient associated with the flexure geometry can be approximated as a constant~\cite{Ugural2003}. Analytical equations for a typical angular hinge based on slender flexures can be used to approximately estimate the relationship between dimensional parameters and the bearing stiffness~\cite{Smith2000, Tielen2014}. Accordingly, the bearing's torsional stiffness can be expressed as 
\begin{equation} \label{eq:k_psi}
K_b^\theta|_{ANA}=\frac{E w {t_p}^3}{12l},
\end{equation}
where $E$ is Young's modulus of the flexure material. $w$, $t_p$, and $l$ are the width, thickness, and length of each leaf, respectively (see Fig. \ref{PivotBearings.ps}(a)).
Under large angles, the pivot bearing material can undergo yielding and plastic deformation. The maximum rotating angle~\cite{Tielen2014} of the bearing can be written as
\begin{equation} \label{eq:theta_max}
\theta_{\text{max}}=\frac{2\sigma_yl}{Et_p},
\end{equation}
where $\sigma_y$ is the yield stress of the flexure material.

In addition to the approximate torsional stiffness Eq. (\ref{eq:k_psi}), we conduct further analysis for the design with more accurate and comprehensive analytical expressions. Specifically, we derive an analytical bearing stiffness matrix using spatial screw theory~\cite{Hopkins2010_FACTP1, Loncaric1987, Bellouard2009, Hopkins2010_Screw}, which contains stiffnesses of the flexure in each of the six DoFs $x,y,z,rx,ry,rz$ with respect to the global frame. Using the dimensions of the various sections of the pivot bearing, the compliance of the pivot bearing in each DoF can be calculated when subjected to a force or torque in any direction. Moreover, the compliance matrix allows the calculation of cross-coupling terms, represented by off-diagonal elements in the stiffness matrix $[K]$, such as the coupling between translation and rotation, which will be used later in Sec. \ref{sec:Dynamic}. In order to minimize the potential interference from other structures and simplify the analysis, the end cap and the PM rotor are assumed to be rigid. The $6\times6$ bearing stiffness matrix with respect to the global (laboratory) frame at the center of the rotational pivot yields
\begin{equation} \label{eq:matrix_stiff}
[K]|_{ANA} = \sum[N^{*}][O]^{-1}[N]^{-1},
\end{equation}
where $[O]$ is the compliance matrix~\cite{Wu2015, Loncaric1987, Ding2004} of each compliant element with respect to its local frame. $[N]$ and $[N^{*}]$ denote the coordinate transformation matrices~\cite{Hopkins2013} from the local to the global frame for the twist and the wrench vectors, respectively. The detailed expressions and derivations of the matrices in Eq. (\ref{eq:matrix_stiff}) are provided in the [supplementary material Sec. S1-3]. Figure \ref{2sizeLSPB.ps}(b),(c) illustrates the analytical BeCu pivot bearing stiffnesses obtained from Eq. (\ref{eq:matrix_stiff}) for several sets of bearing geometries.

\subsubsection{Simulated (FEA) Stiffness Relationships}
Furthermore, an FEA is conducted using the commercial software COMSOL to determine the static stiffness of the pivot bearing. The CAD files of the rotor assembly, including one cylindrical PM connected by two pivot bearings (see Fig. \ref{PivotBearings.ps}(d)), are imported into the COMSOL solid mechanics module for analysis. The rotational and lateral stiffnesses of the bearing are obtained indirectly through the eigenfrequencies, assuming that each eigenmode is predominantly characterized by either purely translation or rotation without coupling. The eigenfrequencies $f_b$ of a mechanical oscillator are proportional to stiffnesses $K_b$. Hence, the rotational and lateral stiffnesses of the oscillator can be expressed as follows

\begin{equation}\label{eq:k_x}
    \begin{cases}
         K_{b}^{\theta} = J(2\pi f_b^{\theta})^2 \\[8pt]
         K_{b}^{l} = M(2\pi f_b^{l})^2 
    \end{cases},
\end{equation}
where $J$ is the moment of inertia of the oscillator around the rotational axis, $M$ is the mass of the oscillator, and $f_b^{\theta}$, $f_b^{l}$ are the eigenfrequencies of the oscillator's rotational and lateral eigenmodes, respectively. 

In the FEA, the flexures of the modeled BeCu pivot bearing are assigned Young’s modulus of 129~GPa and Poisson's ratio of 0.3. In addition, it is assumed both the end cap and the PM rotor are rigid. We record both rotational and lateral eigenmodes through the eigenfrequency analysis in COMSOL and subsequently apply a stiffness conversion using Eq. (\ref{eq:k_x}). Simulation-converted rotor assembly stiffnesses corresponding to various sets of bearing geometries are depicted in Fig. \ref{2sizeLSPB.ps}(b),(c) as solid data points. A power-law fitting (see [supplementary material Sec. S4]) of the FEA results with respect to the flexure dimension parameters is also illustrated in the plot, demonstrating a very good agreement with the analytical curve. 

Fig. \ref{2sizeLSPB.ps}(b) shows the plot of lateral stiffness versus flexure length $l$. The lateral stiffness of the pivot bearing increases approximately 10 times as the flexure length $l$ changes from 20~mm to 10~mm while the other dimensions remain the same (see Fig. \ref{2sizeLSPB.ps}(a)). This value is comparable to the expected $2^3=8$ change which is analytically predicted for simple flexure beams~\cite{Smith2000}.

\begin{figure}[htb]
\centerline{\psfig{figure=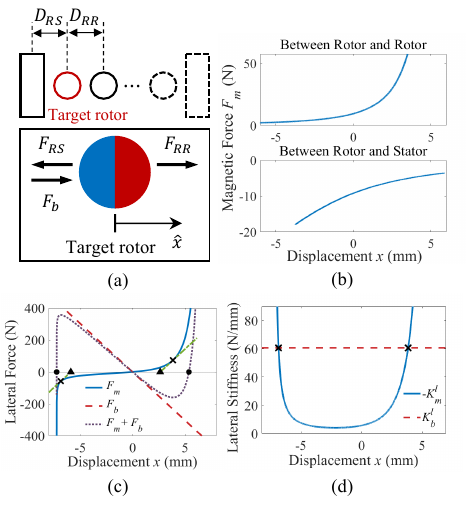,width=0.5\textwidth}}
\caption{Stability of multi-rotor pivot bearing MMR. (a) Top: Schematic of multi-rotor pivot bearing MMR, with rectangular shapes representing stators and circular shapes representing rotors. The target rotor is the red circle in the figure, which assumes it is only affected by the adjacent PM on each side. Bottom: Free-body diagram of the target rotor, where $F_{RS}$, $F_{RR}$, $F_{b}$ denote force between stator and rotor, between rotor and rotor, from bearings, respectively. (b) Lateral magnetic forces (power law fitting from the experimental measurements) Vs. displacement of the target rotor, where $D_{RR}$ = 8~mm and $D_{RS}$ = 7.2~mm. Top: From 4~mm diameter × 63.5~mm long rotor, $F_{m} = F_{RR} = 6734/(D_{RR} - x)^{3.12}$. Bottom: From a 1/8 × 1/2 × 3 $\text{in}^3$ (3.175 × 12.7 × 76.2 $\text{mm}^3$) stator, $F_{m} = F_{RS} = -76.62/(x + D_{RS})^{0.2457} + 36.88$. (c) The total lateral forces Vs. displacement of the target rotor, where $F_{m} = F_{RS} + F_{RR}$ is the magnetic force and $F_{b} = -60.64x$ is the bearing force (from 10~mm length BeCu pivot bearings). The plot illustrates the locations of zero total force $F_{m} + F_{b} = 0$ (black dots). The two green dash lines (negatively sloped from the bearing force curve) are tangent to the magnetic force curve at the two intersections (black crosses). The region between the two projection points (black triangles) represents the safe operating range of this example. (d) The slope magnitude of the force-displacement curve in Fig. \ref{Stability.ps}(c) $-K_{m}^{l}$ = $\partial F_m/\partial x$ is the magnitude of magnetic lateral stiffness. $K_{b}^{l}$ is the pivot bearing mechanical lateral stiffness. The intersect points (black crosses) represent the positions where $K_{m}^{l} + K_{b}^{l} = 0$.} 

\label{Stability.ps}
\end{figure}

\subsection{Static Analysis: Device Static Stability}\label{sec:stability}

A critical aspect of designing MMR using pivot bearings is to consider the stability of the rotor under the action of magnetic forces. This effect must be accounted for in the pivot bearing design when using multi-rotor MMR where rotors are placed close to one another. A multi-rotor design can effectively increase the transmitter's magnetic field output while maintaining a small MoI for each rotor, leading to a high resonance frequency for the same total PM volume. Moreover, according to Eq. (\ref{eq:f_theta}), the frequency increases as the distance between magnets decreases. This trend also holds qualitatively between rotors for the multi-rotor MMR, as each PM rotor also contributes magnetic flux to the adjacent rotors~\cite{Jing2023}.  Figure \ref{2rotorSchematic.ps}(a) shows the schematic of the two-rotor MMR system. We note that a magnet placed between two other magnets, where all of the dipoles are in the same orientation parallel to the array, suffers from an unstable equilibrium even in the most ideal scenario. Intuitively, if the middle magnet is displaced towards one of the two side magnets, it would be attracted by an unbalanced force to the closer magnet and will eventually snap to that magnet.  Moreover, in the proposed MMR configuration, the rotor and stators do not have the same size and geometry. Hence, in addition to the unstable equilibrium described above, each rotor experiences an asymmetric magnetic field from adjacent PMs in the multi-rotor MMR system. This asymmetry can magnify the loss of equilibrium resulting from the unbalanced magnetic forces exerted on each rotor, thereby causing the deviation of rotors in the lateral direction. This section describes the stability consideration when designing a multi-rotor MMR supported by pivot bearings.

We start by examining and evaluating the magnetic forces acting on the rotor. A well-designed MMR should position the rotor at an equilibrium point where the net magnetic force is equal to zero. To analyze the target rotor, assuming that the only magnetic field is that from the closest PM on each side, two magnetic force-distance measurements are needed. As an example, we examine the target rotor shown in Fig. \ref{Stability.ps}(a), where the closest PM on its left ($-\hat{x}$ direction) is a stator, and on its right ($+\hat{x}$ direction) is another rotor. Fig. \ref{Stability.ps}(b) shows the lateral magnetic force-displacement results (fitting curve of the measurement) between two rotors and, as a separate case, between the rotor and the stator. Each RS distance $D_{RS}$ in the plot will have a corresponding RR distance $D_{RR}$ with the same force magnitude. These relations guide the adjustments to $D_{RR}$ and $D_{RS}$ to balance the lateral magnetic force on each rotor.

Next, we specify the stiffness of the bearing $K_{b}^{l}$ to ensure stable operation of the rotor, even when displaced laterally towards the other rotor. We first construct the magnetic force-displacement plot shown in Fig. \ref{Stability.ps}(c), which determines the resultant magnetic force $F_{m}$ on the target rotor at a particular displacement from its origin at the MMR. This magnetic force-displacement plot could be constructed from the addition of the two force-distance plots in Fig. \ref{Stability.ps}(b). In the same figure, we plot the bearing force-displacement relation based on the specified $K_{b}^{l}$ to compare it with the magnetic force. Then, we plot the superposition of the magnetic and bearing forces to show the total force $F = F_{m} + F_{b}$. When the signs of the total force and the displacement are opposite, it indicates a stable restoring force with positive stiffness. In Fig. \ref{Stability.ps}(c), the zero net force positions (indicated by black dots) are marked, with the region in between satisfying the specified condition. This will be the case for forces $F$ in the second and fourth quadrants of the total force curve. 

Theoretically, the assembly is stable as long as the restoring bearing force is larger than or equal to the magnetic force. However, based on experimental evidence, the theoretical condition is not reliable due to the narrow stability range compared to assembly errors and sometimes large lateral deformations due to gravity. Hence, we developed two additional conditions to ensure safe operation within the stability range for any displacement. The first condition is to define the range where the absolute value of the bearing force gradient with respect to distance $F_{b}$ is greater than or equal to the gradient of the magnetic force $F_{m}$. Accordingly, the stiffness-displacement plot shown in Fig. \ref{Stability.ps}(d) is obtained from the gradient of the force-displacement plot. The plot identifies two intersection points between the pivot bearing mechanical stiffness $K_{b}^{l}$ and the magnetic stiffness $-K_{m}^{l}$ curves. These two intersections, indicated by black crosses in Fig. \ref{Stability.ps}(c), represent the positions where the tangent lines of the $F_{m}$ curve have the same magnitude of the slope as the $F_{b}$ curve. The second condition sets a tighter constraint on the safe operation to ensure stability even when the assembly is inaccurate or uncertain. In practice, misalignment errors during assembling or fabrication processes are inevitable. These unpredictable errors shift the force-displacement curve from its origin, leading to an initial force imbalance. To set the second condition, we mark the intersections (black triangles) between the tangent lines and the zero-force axis. A safer operating range is achieved if the target rotor is constricted between these two intersections. We used this condition to set the spacings in the MMR, and it eliminated the instabilities between the rotors and stators during assembly.

\subsection{Dynamic Analysis}\label{sec:Dynamic}

In this section, we analyze the equations of motion governing the PM rotor in the single-rotor MMR with pivot bearings. Importantly, in addition to the fundamental rotational mode based on which the MMR concept is built, experimental evidences underscore the existance of the undesirable transverse (also referred to as lateral) eigenmode in the pivot bearing MMR. Hence, we present a model that not only captures the simplified rotational mode, but also the transverse mode as well as the interesting frequency jumps associated with the mode-crossing effects observed in the experiments.  

\subsubsection{Eigenfrequency of MMR without Pivot Bearings}

In the previous static analysis of Sec. \ref{sec:rotorstatic}, we derived expressions for the static magnetic reaction force and torque on the PM rotor. According to Eq. (\ref{eq:magnetic_stiffness}), under the assumption of linear behavior at low driving amplitudes, the resonance frequency of an MMR considering only the effects of the magnetic component can be estimated as
\begin{equation} \label{eq:f_theta}
f_m^{\theta} = \frac{1}{2\pi} \sqrt{\frac{K_m^{\theta}}{J}}=\sqrt{\frac { \mu_0 m_R m_S}{ 4\pi^3 {D_{RS}}^3 J}},
\end{equation}
where $J$ is the moment of inertia of the oscillating components. In the case of large angles and pure rotation ($x = 0$), we can apply nonlinear analysis to Eq. (\ref{eq:magnetic_load_taylor}). By ignoring higher-order terms beyond the third order, the multiple-scale method~\cite{Vakakis1993} is used to predict the resulting resonance nonlinear backbone curve as a function of the maximum response angle $\theta_{\text{res}}$ as~\cite{Thanalakshme2022}
\begin{equation}\label{eq:f_theta_nonlinear}
f_{m, \text{nonlin}}^{\theta} = f_m^{\theta} - \frac{K_{m}^{\theta}}{64\pi^2 J f_m^{\theta}} {\theta_{\text{res}}}^2.
\end{equation}
As the oscillating amplitude ${\theta_{\text{res}}}$ increases, the system's resonance frequency decreases, leading to a nonlinear softening effect.

However, the translational magnetic stiffness $K_{m}^{l}$ is a negative value, indicating the absence of a translational resonance frequency from the magnetic component and instead an instability. This inherent instability of MMR in the lateral direction induced by the PMs highlights the necessity of mechanical bearings possessing translational stiffness in a practical MMR system. With the additional positive stiffnesses provided by the pivot bearings, as discussed in Sec. \ref{sec:bearingstatic}, the system is now stable and can be operated at the designed resonance frequency.

\begin{figure}[htb]
\centerline{\psfig{figure=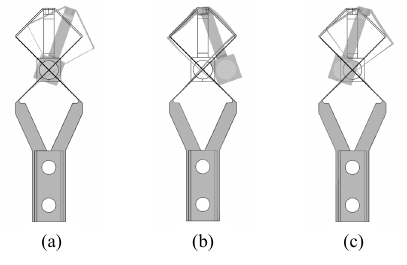,width=0.45\textwidth}}
\caption{Orthogonal (top) view along the pivot bearing axis ($l$ = 20~mm) rotor assembly's eigenmodes (a deformed shape with the superimposed undeformed shape) from COMSOL simulation (a) Rotational eigenmode, (b) Lateral eigenmode, (c) Hybrid eigenmode.}
\label{ModeShape.ps}
\end{figure}

\subsubsection{Equations of Motion and Eigenfrequencies of MMR with Pivot Bearings}

The schematic in Fig. \ref{ModeShape.ps}(a),(b) illustrates a top view of the rotational and lateral eigenmode for the pivot bearing rotor in a single-rotor MMR. Given that the pivot bearing single-rotor MMR system (see Fig. \ref{PivotBearings.ps}(d)) is primarily influenced by its rotational and lateral motions, we can reduce the stiffness matrix $[K]$ in Eq. (\ref{eq:matrix_stiff}) to a $2\times2$ stiffness matrix with only rotational and lateral DoFs. Consequently, the free dynamic response of the single-rotor MMR system can be expressed as

\begin{equation}\label{eq:dynamic_eom_main}
\left[\begin{matrix}J&0\\0&M\\\end{matrix}\right]\left[\begin{matrix}\ddot{\theta}\\\ddot{x}\\\end{matrix}\right]+\left[\begin{matrix}C\\\end{matrix}\right]_{2\times2}\left[\begin{matrix}\dot{\theta}\\\dot{x}\\\end{matrix}\right]+\left[\begin{matrix}K_b^\theta&K_b^{cp}\\K_b^{cp}&K_b^l\\\end{matrix}\right]\left[\begin{matrix}\theta\\x\\\end{matrix}\right]-\left[\begin{matrix}T_m\\F_m\\\end{matrix}\right]=0,
\end{equation}
where $K_b^{cp}$ is the mechanical coupling term between the rotational and lateral DoFs induced by the pivot bearing structure, and has units of N/rad. The $2\times2$ matrix $\left[\begin{matrix}C\\\end{matrix}\right]_{2\times2}$ represents the system's overall damping coefficient. Note that given a positive displacement, the magnetic force $F_m$ and torque $T_m$ expressed in Eq.(\ref{eq:magnetic_load_taylor}) are positive (destabilizing) and negative (restoring), respectively.

In the case of an undamped system with small operating amplitudes, we substitute Eq. (\ref{eq:magnetic_stiffness}) to Eq. (\ref{eq:dynamic_eom_main}) by neglecting high-order terms of the magnetic loads. Subsequently, the equation of motion can be simplified to a normalized linear form

\begin{equation}\label{eq:dynamic_eom_linear}
\left[\begin{matrix}\ddot{\theta}\\\ddot{x}\\\end{matrix}\right]+\left[\begin{matrix}J&0\\0&M\\\end{matrix}\right]^{-1}\left[\begin{matrix}K_b^\theta+K_m^\theta&K_b^{cp}\\K_b^{cp}&K_b^l+K_m^l\\\end{matrix}\right]\left[\begin{matrix}\theta\\x\\\end{matrix}\right]=0.
\end{equation}
Solving the characteristic equation of Eq. (\ref{eq:dynamic_eom_linear}) yields the natural frequencies $\omega_n$ of the coupled system 

\begin{equation}\label{eq:natural_freq}
{\omega_n}^2=\frac{1}{2}\left({\omega_\theta}^2+{\omega_l}^2\pm\sqrt{\frac{4}{JM}\left({K_b^{cp}}\right)^2+\left({\omega_\theta}^2-{\omega_l}^2\right)^2}\right),
\end{equation}
 where $\omega_\theta = \sqrt{\frac{K_b^\theta+K_m^\theta}{J}}$ and $\omega_l = \sqrt{\frac{K_b^l+K_m^l}{M}}$ are the rotational and lateral natural frequencies of a system without DoF coupling effect (i.e., $K_b^{cp} = 0$). From the equation, it is evident that the values of the two natural frequencies $\omega_n$ always lie beyond the range defined by the two uncoupled frequencies, $\omega_{1} < \{\omega_\theta,~\omega_l\} < \omega_{2}$, due to the presence of the coupling term $K_b^{cp}$. 
 The corresponding eigenvectors of Eq. (\ref{eq:natural_freq}) can be expressed as
 
\begin{equation}\label{eq:eigenvectors}
v=\left[{\omega_\theta}^2-{\omega_l}^2\pm\sqrt{\frac{4}{JM}\left({K_b^{cp}}\right)^2+\left({\omega_\theta}^2-{\omega_l}^2\right)^2},\ \ 2\frac{K_b^{cp}}{M}\right]^T.
\end{equation}

\subsubsection{Mode-crossing: Rotational Versus Lateral Oscillations}\label{sec:mode_crossing}

Depending on the induced magnetic field $B_S$ from the stators, the single-rotor MMR demonstrates an interesting interplay between the desired torsional mode and the undesired lateral mode. In principle, alternating magnetic signals are transmitted from the angular oscillations of the magnetic rotor, while the lateral modes dissipate power, through viscous damping, without propagating useful alternating magnetic flux. According to Eq. (\ref{eq:magnetic_stiffness}), the rotational and lateral magnetic stiffnesses of the system are both functions of the field $B_S$, indicating that changing $B_S$, or equivalently adjusting $D_{RS}$, will affect the magnitudes of $\omega_\theta$ and $\omega_l$ in our system. When $\omega_\theta = \omega_l$, Eq. (\ref{eq:natural_freq}) and Eq. (\ref{eq:eigenvectors}) indicate that the coupling $K_b^{cp}$ has the greatest influence on the system natural frequency and mode shape. As the value of $\omega_\theta$ moves away from $\omega_l$, the impact of $K_b^{cp}$ on the system becomes comparably weaker, with the system's eigenfrequencies $\omega_n$ approaching $\omega_\theta$ and $\omega_l$. This suggests that to maximize rotational motion at resonance, we should design an MMR system with $\omega_\theta$ far away from $\omega_l$.

Table \ref{TableModeCrossing} lists the approximations of eigenfrequencies of the pivot bearing single-rotor MMR under different scenarios, which will be introduced subsequently. This table can be used for quick design of the device and for validation of parameter selections.
First, we only consider a single-rotor pivot bearing MMR without stators, where the stiffness is purely mechanical from the pivot bearings. In this configuration, the coupling term has a minimal effect on the system. According to our design, the eigenfrequency of the lateral mode shape is greater than that of the torsional mode, and therefore can be approximated as $\sqrt{K_{b}^{l}/M}$ and $\sqrt{K_{b}^{\theta}/J}$, respectively. Next, when adding stators to the system, the magnetic torsional stiffness $K_{m}^{\theta}$ is directly proportional to the magnetic field $B_S$ produced by the stators, as shown in Eq. (\ref{eq:magnetic_stiffness}). Consequently, the device eigenfrequency of the rotational mode increases with the increase in $B_S$. However, at the same time, the eigenfrequency of the lateral mode may not have a huge variation up to a certain $B_S$ due to $K_{b}^{l} \gg |K_{m}^{l}|$ for low $B_S$. When the device angular frequency $\omega_{\theta}$ approaches the magnitude of lateral frequency $\omega_{l}$, the two modes start to interact with each other, forming two hybrid modes that consist of both rotational and lateral motions as demonstrated in Eq. (\ref{eq:eigenvectors}). Figure \ref{ModeShape.ps}(c) shows a hybrid mode shape existing in the pivot bearing single-rotor MMR.

\begin{table}[htb]
\begin{center}
\caption{Approximated resonance frequencies (undamped) of the different bearing-magnet configurations (arranged in increasing magnitude of $B_S$ from top to bottom). This table can be used for quick design of the device and for validation of parameter selections. The level of $B_S$ required to reach the hybrid regime depends on the pivot bearing structure. In the hybrid regime, Eq. (\ref{eq:natural_freq}) is required to determine the system resonance frequency as the coupling term $K_b^{cp}$ has a relevant great effect on the system's resonance frequency.}
\label{TableModeCrossing}
\begin{tabular}{c M{1.8cm} M{1.8cm}}
& & \\ 
\hline
\hline \\[-2ex]
Configuration & $\omega_{1}$ & $\omega_{2}$ \\ [1ex] 
\hline \\[-1ex]
Bearing Only & $\sqrt{K_{b}^{\theta}/J}$ & $\sqrt{K_{b}^{l}/M}$ \\ [1ex]
Before Hybrid Regime & $\omega_{\theta}$ & $\omega_{l}$ \\ [1ex]
Hybrid Regime & \multicolumn{2}{c}{ See Eq. (\ref{eq:natural_freq}) } \\ [1ex]

After Hybrid Regime& $\omega_{l}$ & $\omega_{\theta}$ \\ [1ex]

\hline
\hline
\end{tabular}
\end{center}
\end{table}

The transition region where the device's rotational and lateral modes interact is called the ``hybrid regime". At the beginning of the hybrid regime, the angular motions dominate in the lower-frequency eigenmode, while the higher-frequency eigenmode has mostly lateral motions. As $B_S$ increases, the two eigenmodes switch identities, meaning that the secondary motions gradually become more apparent and eventually take precedence over the primary motion. At the end of the hybrid regime, the lower-frequency eigenmode is mainly lateral, while the angular motions dominate the higher-frequency eigenmode. We term this dominating mode transition circumstance as ``mode-crossing", which happens at $\omega_{\theta} = \omega_{l}$. Importantly, mode interactions significantly affect the power efficiency of the MMR. In the hybrid regime, the lateral motion consumes part of the energy used to generate rotational oscillation. Those energies are eventually damped and wasted in the form of heat.

\section{Experimental Methods and Materials}\label{sec:Experimental}
To characterize the behavior of the pivot bearing when used in an MMR system, two key performance indicators are studied: the effective lateral stiffness and the device damping coefficient. Four different types of experimental measurements are conducted to obtain the performance indicators: (1) frequency sweep test, (2) force-displacement test, (3) tuning fork free-response test, and (4) single-rotor device free-response test. From the frequency sweep test, we measure the system resonance frequencies and the overall damping coefficient. 
A force-displacement test performed with an Instron characterizes the magnetic forces between PMs, which is used to conduct the stability analysis. The tuning fork free-response test compares the intrinsic damping among the candidate materials, to determine the most energy-conserving material for fabricating the pivot bearings. The single-rotor device free-response test obtains the damping coefficient of our single-rotor pivot bearing MMRs. [supplementary material Sec. S8] provides further information for our experimental setups.
\subsection{Frequency Sweep Test Setup}
The frequency sweep test of pivot bearing MMR is used to measure the resonance frequency and the power efficiency of the system.  
Additionally, the softening nonlinearity shown in Eq. \ref{eq:f_theta_nonlinear} is captured in the frequency sweep test by changing the driving voltage amplitude. The electromagnetic flux excitation is generated by a 100-turn coil (95~mm × 135~mm inner dimensions) made of AWG 18 enameled copper wire, placed next to the MMR. An HP 33120A function generator controls the frequency and amplitude of the driving coil's voltage. A Gemini XGA-5000 power amplifier connects the function generator and the coil, boosting the current flowing into the circuit. 
The combined output magnetic flux density, comprising the field generated by the MMR and the driving coil, is tracked by a Hall effect sensor positioned 10~in (0.254~m) horizontally from the device. A Hantek CC-65 current clamp measures the current flow through the coil. The voltage across the coil is measured from a Tektronix TPPO200 voltage probe. A Tektronix DPO 2014B digital phosphor oscilloscope is used to monitor the magnitude of coil current, voltage, and magnetic flux. The system power dissipation is calculated by multiplying the instantaneous values of coil current and voltage. Two Stanford Research Systems SR860 lock-in amplifiers are used to record the root mean square (RMS) value of the magnetic flux density as well as the RMS value of the current flow through the coil at each frequency. The lock-in amplifiers also measure the phase differences among the current flow through the coil, the magnetic flux signal received at the sensor, and the voltage output from the function generator. 
\subsection{Magnetic Force-displacement Test Setup}

We use an Instron universal testing machine to measure the force between two PMs as a function of the distance between them. We fasten two PMs onto the holders and orient them along the same polarization direction. The magnets are initially brought in contact and then separated by the machine at a constant velocity. The lateral forces as a function of separation distance are collected and used in the device stability analysis.

\subsection{Tuning Fork Free-response (Ring-down) Test Setup} \label{sec:tuningforksetup}

The tuning fork free-response tests quantify the material's intrinsic damping. In general, without a consistent external excitation, the damping force will keep attenuating the oscillator until it rests. However, the system damping loss originates from not only the material's intrinsic damping but also some extrinsic dissipation, such as contact friction dissipation and anchor loss. 

To exclusively quantify the material's intrinsic damping, minimizing the effect of other losses, we compare the dissipation of fabricated tuning forks made from each material. While prongs oscillate, the symmetrical structure of the tuning fork cancels out the lateral motion at its handle, thus minimizing the extrinsic dissipation during vibration due to anchor losses. The tuning fork handle is fixed by a metal clamp, which is positioned at a distance of 15~mm from the handle's bottom end.

We apply an impulse excitation by gently hitting the top of one prong with a wooden stick. A microphone placed next to the tuning fork prong measures the amplitude of the free-response decay acoustic wave. 
The Q-factors are calculated by fitting the exponentially decaying envelope from each measurement. Q-factors are used to quantitatively compare the intrinsic damping of different materials at the resonance frequency.

\subsection{Material Selection using Tuning Fork Ring-down Tests}

\begin{figure}[htb]
\centerline{\psfig{figure=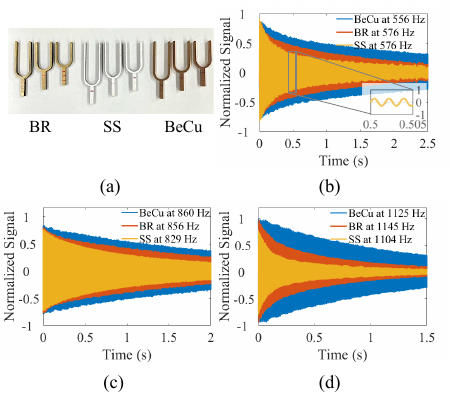,width=0.5\textwidth}}
\caption{ (a) Tuning forks of pitches around 500~Hz, 750~Hz, and 1000~Hz, left to right respectively, from brass (left three), stainless steel (middle three), and copper beryllium (right three). (b-d) Tuning fork ring-down test oscillation decay curves of (b) 500~Hz, (c) 750~Hz, and (d) 1000~Hz. The normalized signal converted from audio represents the amplitude of tuning fork oscillation. Inset of (b): magnified view showing the detailed oscillation curve from $t$ = 0.5 to 0.505~s. Comparing the exponential decay envelopes among materials at each frequency, we can qualitatively conclude that BeCu tuning forks have the largest Q-factor hence the smallest intrinsic damping.}
\label{TuningFork.ps}
\end{figure}

Reducing the material intrinsic loss in the pivot bearing improves MMR's power efficiency to generate magnetic signals. From the Ashby chart of the material resilience and loss coefficient~\cite{ASHBY2006}, we select SS, Al, BR, and BeCu as our candidate materials used for the pivot bearing. In order to validate their intrinsic damping and machinability, we manufacture tuning forks from BR, SS, and BeCu and conduct ring-down tests as we mention in Sec. \ref{sec:tuningforksetup}. The different material properties, such as the density and Young’s modulus, affect the eigenfrequency of the tuning fork. We set the length of the two prongs as a parameter in the Abaqus FEA and proceed with a parametric sweep to find the corresponding eigenfrequencies of the forks' out-of-phase mode. Then, we make adjustments to the tuning fork dimensions based on the FEA results for different materials, guaranteeing that they can produce approximately the same pitches. Figure \ref{TuningFork.ps}(a) shows the tuning fork specimens made out of BR, SS, and BeCu, for three different pitches.

Figure \ref{TuningFork.ps}(b)-(d) show the ring-down test results comparison among the three tuning fork materials at different pitches. Based on the decay envelope and Q-factor comparison, BeCu tuning fork has the lowest power dissipation, thus is the best material used for our pivot bearing fabrication.

\subsection{Pivot Bearing Single-rotor Device Free-response (Ring-down) Test Setup}

The free-response test of the pivot bearing MMR is designed to measure the damping coefficient of the pivot bearing single-rotor MMR at its rotational resonance frequency. The MMR is mounted onto an acrylic substrate using screws. The device's damping coefficient causes a gradual decay of the oscillation, from an initial angle at the beginning of the test to rest when all of the energy has been dissipated. A unit-pulse magnetic flux excitation around MMR's resonance frequency is applied to the MMR by a coil to excite the rotor's initial angular oscillation. A Hall effect sensor detects the magnetic flux magnitudes generated by the rotor, which can be further mapped to the rotor rotating angles $\theta$. Like the tuning fork free-response test, we compute the Q-factor of the MMR device by fitting the exponentially decaying envelope of the rotor rotation angle. This Q-factor takes into account any additional damping effects that may occur within the device, in addition to the intrinsic damping of the bearing flexure material.
\subsection{Bearing Dimensions and Material Properties}
Table \ref{TableMaterial} lists the mechanical properties of the candidate pivot bearing materials. Equation  (\ref{eq:theta_max}) is used to determine the dimensions of the flexures, ensuring that each pivot bearing has a maximum rotating angle of more than 30 degrees along with a safety factor of 2~\cite{Smith2000}. Table \ref{TableDimension} lists the flexure materials and dimensions for four different types of pivot bearings that have been fabricated.

\begin{table}[htb]
\caption{Pivot bearing material mechanical properties.}
\begin{center}
\label{TableMaterial}
\begin{tabular}{c M{2.5cm} M{2.5cm}}
& & \\ 
\hline
\hline \\[-2ex]
Material & Young's Modulus $E$ (GPa) & Yield Stress $\sigma_y$ (MPa) \\
\hline \\[-2ex]
SS & 190 & 950  \\
Al & 70 & 280  \\
BeCu & 129 & 1000 \\
\hline
\hline
\end{tabular}
\end{center}
\end{table}

\begin{table}[htb]
\caption{Type of fabricated pivot bearings and dimensions.}
\begin{center}
\label{TableDimension}
\begin{tabular}{M{2.4cm} M{0.6cm} M{0.6cm} M{0.6cm} M{0.6cm} M{0.6cm}}
& & \\ 
\hline
\hline \\[-2ex]
Bearing Flexure Material & $l$ (mm) & $s$ (mm) & $t_p$ (mm) & $t_s$ (mm)& $w$ (mm)  \\
\hline \\[-2ex]
SS, Al, BeCu & 20 & 8.33 & 0.1 & 0.3 & 6.35 \\

BeCu & 10 & 3.46 & 0.1 & 0.3 & 6.35  \\
\hline
\hline
\end{tabular}
\end{center}
\end{table}

\section{Results and Discussion}\label{sec:Results}

\begin{figure}[ht]
\centerline{\psfig{figure=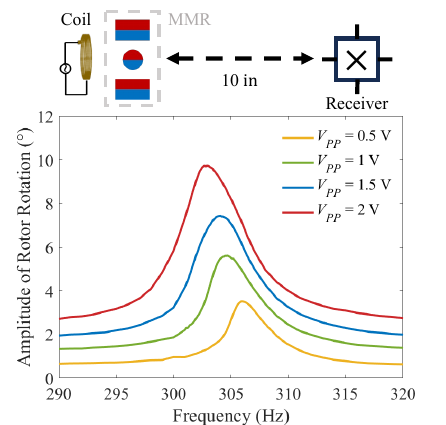,width=0.45\textwidth}}
 \caption{Experimental single-rotor BeCu pivot bearing ($l$ = 10~mm, $t_{e}$ = 2~mm, 12.7 × 12.7 × 76.2 $\text{mm}^3$ stator, $D_{RS}$ = 9.66~mm) MMR backward sweep: rotor rotational amplitude Vs. driving frequency. The rotor's rotational angle is derived from magnetic field measurements obtained from a receiver placed 10 inches away, as illustrated in the schematic at the top. Each distinct colored line on the plot corresponds to a different voltage input amplitude applied to the driving coil. At the highest voltage input $V_{PP} = 2$~V, the device exhibits a damping coefficient of $1.5\cdot 10^{-6}$ ~$\text{N}\cdot\text{m}\cdot\text{s}/\text{rad}$ at its resonance frequency (303~Hz). The nonlinear softening effect is shown on the curve: As the driving amplitude increases, the resonance frequency decreases.}
\label{FreqSweep.ps}
\end{figure}

\subsection{Single-rotor MMR Eigenfrequencies and Eigenmodes} 
The MMR eigenfrequencies (resonance frequencies) are identified through the frequency spectrum plot obtained from the frequency sweep tests. The device exhibits a larger oscillating amplitudes at resonances than at other frequencies. Figure \ref{FreqSweep.ps} shows the measured rotor rotation angle (calculated from the measured field) against the driving frequency for the single-rotor BeCu pivot bearing MMR. Different coil driving amplitudes are shown in the plot with different colors. Due to the softening nonlinear effect represented in Eq. (\ref{eq:f_theta_nonlinear}), the resonance frequencies of the system decrease from 306~Hz to 303~Hz as the driving amplitude increases from 0.5~V to 2~V. This experiment achieves a maximum rotor oscillating angle of approximately 10 degrees with the input of peak-to-peak voltage $V_{PP}$ = 2~V. Overall, the results show stable, smooth rotational oscillations with a weak nonlinear softening effect.

\begin{figure}[htb]
\centerline{\psfig{figure=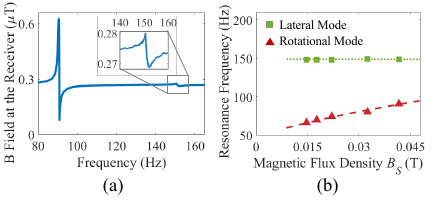,width=0.5\textwidth}}
\caption{(a) Frequency sweep: measured total magnetic field (B field) at the receiver (positioned 10 inches away from the device) Vs. driving frequency of BeCu pivot bearing single-rotor device with 0.042~T external flux density, showing rotational resonance at 90~Hz and lateral resonance at 151~Hz. Inset: magnified view showing the resonance frequency at 151~Hz (b) Experimental results of BeCu pivot bearing ($l$ = 20~mm, $t_{e}$ = 1~mm) single-rotor MMR resonance frequencies Vs. Estimated external magnetic field $B_S$. Dashed lines represent the fitted trend curves for each mode using the fit function $\omega_\theta = 2\pi\sqrt{a^\theta B_S + {f_b^\theta}^2}$ with a fitting parameter $a^\theta$ = $1.4\cdot10^5$ and $\omega_l = 2\pi\sqrt{a^l {B_S}^{5/3} + {f_b^l}^2}$ with $a^l$ = $-1.1\cdot10^4$, where $f_b^\theta = 48~\text{Hz}$ and $f_b^l = 149~\text{Hz}$ are rotational and lateral mechanical resonance of device without stators ($B_S = 0$), respectively. The different external magnetic flux densities $B_S$ that the rotor received are estimated from the FEA in COMSOL with different RS distances.}
\label{2peaks.ps}
\end{figure}

Since the alternating magnetic signals are generated by the rotation of the rotor dipole direction, only the torsional eigenmode is considered useful for signal transmission in MMR devices. However, the rotor lateral eigenmodes and eigenfrequencies which result from the lateral compliance of the flexures must also be considered. The lateral dynamics convey important information regarding the system stability which we have mentioned in Sec. \ref{sec:stability}. Moreover, in this section, we present the dynamics arising from the lateral mode. Firstly, we observe that the magnetic field generated from the lateral eigenmode is considerably small compared with that from the torsional mode, but it can still be detected by a Hall effect sensor. Figure \ref{2peaks.ps}(a) shows the frequency response plot for the 20~mm length BeCu single-rotor magneto-mechanical transmitter system (including the field generated by the driving coil). The torsional resonant frequency appears at 90~Hz on the left-hand side of the plot, while the lateral eigenfrequency appears at 151~Hz on the right-hand side of the plot. Figure \ref{2peaks.ps}(b) shows the MMR resonance frequencies of both eigenmodes at different estimated external flux densities $B_S$ (produced by stators on the rotor), achieved by different $D_{RS}$. $B_S$ are calculated using FEA in COMSOL (details in [supplementary material Sec. S5]), by substituting the experimental PM properties and $D_{RS}$. As we mentioned in the dynamic analysis Sec. \ref{sec:Dynamic} and Table \ref{TableModeCrossing}, the resonance frequencies of our system outside the hybrid regime can be estimated using $\omega_\theta$ and $\omega_l$. Since $K_{b}^{l}$ is much greater than $K_{m}^{l}$ in this MMR setup, the lateral resonances do not vary with the range of different tested external magnetic flux densities. While the torsional resonance frequency varies from 60~Hz to 90~Hz as $B_S$ changes from 0.015~T to 0.042~T. 

\begin{figure}[htb]
\centerline{\psfig{figure=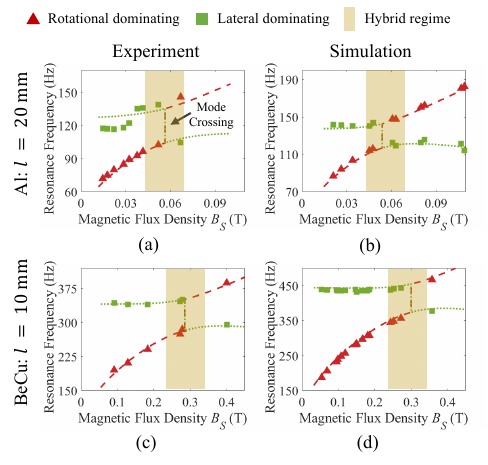,width=0.5\textwidth}}
\caption{Device eigenfrequencies Vs. estimated external magnetic flux density: Al pivot bearing ($l$ = 20~mm, $t_{e}$ = 1~mm) single-rotor MMR from (a) experiment and (b) COMSOL simulation; BeCu pivot bearing ($l$ = 10~mm, $t_{e}$ = 1~mm) single-rotor MMR from (c) experiment and (d) COMSOL simulation with fitting lines (dashed) based on Eq. (\ref{eq:natural_freq}). The fitting parameters for each case can be found in [supplementary material Sec. S7]. The switch between dominating modes (mode-crossing) happens at the intersection of the two fitting lines, where $\omega_\theta = \omega_l$. Within the hybrid regime, mode interactions (hybrid modes) are observed in both experiments and simulations. While visually pure rotational or lateral modes are observed outside the regime.}
\label{ModeCrossing.ps}
\end{figure}

Figure \ref{ModeCrossing.ps} shows the interesting dynamics observed in the resonance measurements of the 20~mm length single-rotor Al pivot bearing MMR and the 10~mm length single-rotor BeCu pivot bearing MMR, and compares them to the COMSOL FEA simulations. Pure rotational and lateral eigenmodes are observed at low and high flux densities. The MMR operation at these regimes is advantageous because the two eigenfrequencies are well-separated and because the eigenmodes are predictable using the simple models presented earlier in this study. In contrast, hybrid mode shapes containing both lateral and rotational motions are recognized within the hybrid regime, which occurs at intermediate flux densities. Within the color-shaded hybrid modes regime, we observe that the lateral frequency drops, and the rotational frequency jumps to a higher frequency, which we refer to as mode-crossing. The simulation solves the eigenfrequencies of the single-rotor pivot bearing MMR by incorporating the solid bearing structure with nonlinear magnetic load in Eq. (\ref{eq:magnetic_load_taylor}) (with further details provided in the [supplementary material Sec. S5]). We observe mode-crossing behaviors in both experiment and simulation, as shown by the data points in Fig. \ref{ModeCrossing.ps}. Using Eq. (\ref{eq:natural_freq}) as a fitting function (dashed lines) for each mode, we perform system identification of the mode-crossing region, and find that the mode-crossing occurs at $\omega_\theta = \omega_l$, consistent with the observed behavior of the devices. A few important insights can be gained from the results of Fig. \ref{ModeCrossing.ps}. First, when a compliant bearing support, like the pivot bearing presented here, is used in magnetically coupled resonators, it is expected that a lateral vibration mode would arise. This unwanted lateral mode would not arise with ball bearing system because of their ultra high translational stiffness. However, even though this might be considered a shortcoming of the use of compliant bearings in MMR, the viscous and friction dissipation of ball bearings and their extremely high pitch acoustic noise generated due to metal ball friction makes them impractical for real-world use. Hence, the pivot bearing studied in this article still provides a considerably higher performance for torsional MMRs. In fact, the effective structural design of the pivot bearing is confirmed by the translational eigenfrequency being higher than the rotational eigenfrequency: it functions as intended where it is compliant in rotation and stiff in translation. Secondly, due to the different proportionality dependence of the magnetic forces and torques on the distances between the magnets, namely $D_{RS}$, it is expected that the order of the frequency would change as a function of these distances and the magnetic field intensity, which is observed in the data when comparing the left to the right-hand sides of the shaded regions of Fig. \ref{ModeCrossing.ps}. Thirdly, the device operation point with the highest data transmission rate is achieved with the 10~mm BeCu bearing operating in rotational mode at a frequency of 380~Hz at a field value of $B_S=0.4$ T. This is a high performance operation regime because the MMR is in static equilibrium, and operates at a high frequency, which is advantageous when MMR are used as data transmitters.

\subsection{Single-rotor Pivot Bearing MMR Damping}
We study the effect of material intrinsic damping on energy dissipation in MMR by fabricating pivot bearing flexures using SS, Al, and BeCu. We assemble each pivot bearing with PMs and accessories to create single-rotor MMRs with the various alloys. Note that BeCu showed the smallest intrinsic damping coefficient through the tuning fork ring-down tests. In this context, the damping coefficient of the MMR system is defined only with respect to the rotational motion of the PM rotor, as effective magnetic signals are generated only from rotation, not lateral motion. We conduct the ring-down experiments for each single-rotor MMR with different $D_{RS}$, from 19.05~mm to 31.75~mm. At the same time, we closely monitor the undesired lateral eigenfrequency, making sure that all MMR testing setups are kept before the hybrid regime as mentioned in Sec. \ref{sec:mode_crossing}. 

When $K_{m}^{\theta}$ is negligible at very large $D_{RS}$, the damping coefficient of the BeCu pivot bearing MMR equals $3.5\cdot 10^{-8}$ N$\mathrm{\cdot}$m$\mathrm{\cdot}$s/rad, which is less than the SS pivot bearing MMR by a factor of 4.6 and less than the Al pivot bearing MMR by a factor of 1.8. This result proves that the BeCu pivot bearing has the lowest intrinsic damping among our bearing selections.

\begin{figure}[htb]
\centerline{\psfig{figure=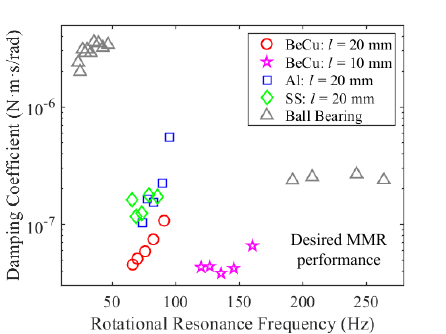,width=0.5\textwidth}}
\caption{Experimental pivot bearing single-rotor MMRs' damping coefficient Vs. device rotational resonance frequency. The damping coefficients are obtained from the ring-down decay envelopes. Color represents the type of pivot bearing used in the MMR. The experimental data points for each device are obtained using stators with dimensions of 3.175 × 12.7 × 76.2 $\text{mm}^3$, while changing MMR's $D_{RS}$ from 1.25~in (31.75~mm) to 0.75~in (19.05~mm), corresponding to $B_S$ from 0.01~T to 0.04~T. A single-rotor ball bearing device with the same PM setups (192~Hz to 264~Hz) and a single-rotor ball bearing device with large PM sizes \cite{Kanj2022} (23~Hz to 47~Hz) are included as references. The desired MMR device should have a low damping coefficient and a high resonance frequency, which is located at the bottom right of the plot.}
\label{DeviceDamping.ps}
\end{figure}

In this case at small $D_{RS}$, both $K_{m}^{\theta}$ and $K_{b}^{\theta}$ contribute to the torsional stiffness of the resonator. Figure \ref{DeviceDamping.ps} shows the ring-down test results comparison of single-rotor MMR with the various pivot bearings at their rotational eigenfrequencies. Data points are obtained using identical experimental setups except for the different pivot bearings, which are indicated by the different colors. Additionally, results from ball bearing single-rotor devices are included for reference. The different resonance frequencies from the same bearing are accomplished by different $K_m^\theta$ resulting from changing $D_{RS}$. Comparing the results from each bearing, we notice that the BeCu pivot bearing MMR always shows the smallest damping among all the MMR configurations at each frequency.

Due to the large MoI of the 20~mm length pivot bearing, it is challenging to drive the MMR device frequency to more than 150~Hz even with very small $D_{RS}$. Therefore, another BeCu pivot bearing with 10~mm length is manufactured and tested (see Fig. \ref{2sizeLSPB.ps}(a)). The damping coefficients from the ring-down of the 10~mm length BeCu pivot bearing single-rotor MMR are also shown in Fig. \ref{DeviceDamping.ps}. This smaller pivot bearing reaches a resonance frequency of 162~Hz at $D_{RS}$ = 19.05 mm. In comparison, the single-rotor devices using ball bearings under the same $B_S$ reach 250~Hz due to the small MoI of the bearings as shown in Fig. \ref{DeviceDamping.ps}.

Besides the challenge of pushing frequencies, the length $l$ of the bearing also impacts the damping coefficient of the device. As discussed in Sec. \ref{sec:mode_crossing}, the resonance mode shape could be affected by the coupling term when operating near the hybrid regime, where $\omega_\theta \approx \omega_l$, resulting in increased undesired lateral motion and consequently greater dissipation. Since reducing flexure length could significantly increase $K_{b}^{l}$ as Fig. \ref{2sizeLSPB.ps}(b), the lateral resonance frequency $\omega_l$ of the 10~mm length BeCu pivot bearing device is pushed to a much higher frequency, approximate 450~Hz (with $t_e$ = 2~mm), which is far away from the operating frequencies. This is in contrast with the operating frequencies for the 20~mm length bearing devices which are relatively close to their $\omega_l$ at 150~Hz. From Fig. \ref{DeviceDamping.ps}, we observe the expected increase in damping as the resonance frequency $\omega_\theta$ increases, and the increase is more noticeable in pivot bearing with a bigger size ($l$ = 20~mm) due to their comparably smaller $\omega_l$ value.

The 10~mm length bearing device shows smallest damping coefficients among all setups with different $D_{RS}$, as presented in Fig. \ref{DeviceDamping.ps}. The damping coefficients of the two BeCu pivot bearing devices with different $l$ at $D_{RS}$ = 31.75~mm are approximately equal, proving the excellent intrinsic damping properties of BeCu pivot bearing at pure rotational mode. Such damping coefficient value is 5 times smaller than that of the device with ball bearings under similar conditions (data points from 192~Hz to 264~Hz in Fig. \ref{DeviceDamping.ps}). Comparing the pivot bearing with the ball bearing MMR \cite{Kanj2022} operating at 23~Hz to 47~Hz (see Fig. \ref{DeviceDamping.ps} top left corner), the proposed pivot bearing device demonstrates an 80-fold reduction in damping coefficient. The high damping coefficient of that ball bearing device comes from the strong lateral magnetic force that the stators apply to the large PM rotor during operation, resulting in increased friction loss.

\subsection{Multi-rotor Interlocked Pivot Bearing MMR}
 
\begin{figure}[htb]
\centerline{\psfig{figure=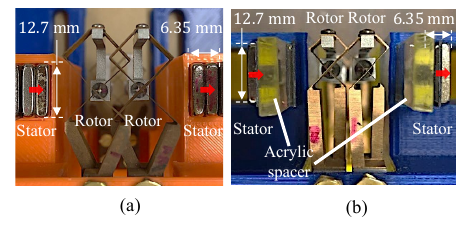,width=0.5\textwidth}}
\caption{ Top view of BeCu pivot bearing two-rotor MMR devices with 6.35 × 12.7 × 76.2 $\text{mm}^3$ stator: (a) $l$ = 20~mm, $t_e$ = 2~mm, $D_{RR}$ = 10~mm, and $D_{RS}$ =  11.5~mm, with device rotational resonance frequency at 113~Hz. (b) $l$ = 10~mm, $t_e$ = 2~mm, $D_{RR}$ = 7~mm, and $D_{RS}$ =  9.175~mm, with device rotational resonance frequency at 285~Hz. The red arrows indicate the polarization direction of the PMs.}
\label{2rotorSity.ps}
\end{figure}

Interlocked design enables compact multi-rotor MMR, where the length $l$ of the pivot bearing flexures is greater than $D_{RR}$. Figure \ref{2rotorSchematic.ps}(b) illustrates the CAD schematic of a two-rotor interlocked MMR design. As we mentioned in Sec. \ref{sec:stability}, rotor position adjustment needs to be implemented to minimize the unbalanced lateral magnetic force. Figure \ref{2rotorSity.ps} shows the top view of the manufactured two-rotor interlocked MMR with BeCu pivot bearings of flexure length $l$ = 20~mm (see Fig. \ref{2rotorSity.ps}(a)) and $l$ = 10~mm (see Fig. \ref{2rotorSity.ps}(b)), which have the rotational resonance frequencies of 113~Hz and 285~Hz, respectively. These two successful setups show the validity and feasibility of this pivot bearing design. However, due to the limitation of the manufacturing process mentioned in Sec. \ref{sec:design_constraints}, the pivot bearing needs a large $l$ to maintain an appropriate maximum angle. This drawback results in a large magnitude of the rotor assembly's MoI, which sacrifices the maximum reachable resonance frequency of the MMR. If we could fabricate a pivot bearing into a smaller size using other precision machinery technology, such as laser micromachining, we would be able to further increase the MMR's resonance frequency.

The stability model described in Sec. \ref{sec:stability} is also validated by constructing several two-rotor pivot bearing MMR devices. 
In the experiment, we observed that the BeCu pivot bearing of length $l$ = 20~mm and end cap thickness $t_{e}$ = 1~mm (see Fig. \ref{PivotBearings.ps}(b)) can not hold the two-rotor MMR from snapping towards each other when the RR distance is 8~mm or less. Qualitatively, we concluded that the Al end cap causes a finite parasitic compliance that needs to be considered in the static analysis described in Sec. \ref{sec:bearingstatic}. Ideally, this end cap should be perfectly rigid without affecting any pivot bearing mechanical properties. However in practice, we made this end cap small and thin to minimize the MoI of the oscillator, causing a finite amount of flexibility in the rotor support. We qualitatively observe that the bearing effective lateral stiffness is proportional to the thickness of the end cap $t_{e}$. 
Using Eq. (\ref{eq:k_x}), we estimate the lateral stiffness provided from the pivot bearing to be 6.5~N/mm, where $f^l$ is 149~Hz and the effective mass $M$ is 7.5~g as measured from the experiment. 
Substituting the pivot bearing properties into the analysis in Sec. \ref{sec:stability}, the safe operating range (second condition) at 8~mm $D_{RR}$ for the target rotor (with a stator on its $-\hat{x}$ side and a rotor on its $+\hat{x}$ side, similar to Fig. \ref{Stability.ps}(a)) is [-1.40~mm, 0.04~mm]. The very small number on the $+\hat{x}$ boundary could explain the reason for failure: any rotor misalignment greater than 0.04~mm to the $+\hat{x}$ will introduce instability. Table \ref{TableStability} lists the safe operating ranges for each pivot bearing device with different RR distances. We can clearly see that the $l =10$~mm pivot bearing device has a much wider safe operating range in the lateral direction than the $l =20$~mm, following the relations of mechanical lateral stiffness shown in Fig. \ref{2sizeLSPB.ps}(b).

\begin{table*}[t]
\caption{The safe operating ranges (in millimeters) of device pull-in stability boundaries for the target rotor of various BeCu pivot bearing two-rotor MMR.}
\begin{center}
\label{TableStability}
\begin{tabular}{c l l l l}
& & \\ 
\hline
\hline \\[-2ex]
Pivot Bearing MMR & $D_{RR}$ = 7~mm & $D_{RR}$ = 8~mm & $D_{RR}$ = 9~mm & $D_{RR}$ = 10~mm \\
\hline \\[-2ex]
$l$ = 20~mm; $t_{e}$ = 1~mm & N/A & [-1.40, 0.04] &[-3.57, 0.44] &[-5.48, 1.08]   \\

$l$ = 20~mm; $t_{e}$ = 2~mm & [-1.21, 0.14] & [-3.56, 0.73] &[-5.85, 1.50] &[-7.80, 2.35]  \\

$l$ = 10~mm; $t_{e}$ = 2~mm & [-3.38, 1.70] & [-5.89, 2.63] &[-8.23, 3.59] & $>$ 4.5  \\
\hline
\hline
\end{tabular}
\end{center}
\end{table*}
 
\section{Conclusion}\label{sec:Conclusion}
 
In this work, we designed and fabricated a new pivot bearing based on the concept of flexure hinges. Magneto-mechanical resonators (MMR) constructed using the pivot bearing produce a time-varying magnetic field with a frequency range from 50~Hz to 306~Hz. The proposed pivot bearing can be designed with large transverse stiffness while maintaining a maximum rotating angle of more than 30 degrees, which is a factor of 2 higher than commercial pivot bearings~\cite{C-Flex2019}. We demonstrate multiple MMR devices and features, including multi-rotor MMR built using the pivot bearing with the interlocked design. We show that BeCu is the optimal choice among our selected materials for pivot bearing fabrication because of its superior material intrinsic damping during oscillation. This BeCu pivot bearing MMR demonstrates a 5 to 80 times lower system damping coefficient compared to the previous ball bearing MMR~\cite{Kanj2022}. Additionally, analytical equations for the stiffness of the pivot bearing are derived and verified through FEA, and static stability analysis —against pull-through snapping— of the MMR with pivot bearing is developed and experimentally validated. Finally, we study the interaction between torsional and lateral oscillation modes which influence the dissipation during the operation of the MMR devices and show how the flexure geometry and the strength of the magnetic field from the stators can be tailored to achieve high resonance frequency and low damping while minimizing the undesired lateral oscillation of the rotor. Notably, we derived an analytical and numerical framework to study the hybrid rotational-translational modes arising at intermediate magnetic fields. These modes are important for all magneto-mechanical oscillators using elastic bearings ranging from the micro to the macro scale, including the wireless communication devices used in medicine~\cite{Gleich2023}. Further improvements could involve topology optimization of the flexure structure, such as shortening the flexure length to increase transverse stiffness by $l^{-3}$ and reduce the bearing MoI by $l^{2}$, or exploring non-homogeneous materials (or metamaterials) to further enhance transverse stiffness. Because we provide experimental data and modeling of this new bearing, we anticipate that future research in this area will optimize the geometry and construction further by established design optimization methods. Overall, these advancements could enable MMR devices with high performance for ultralow-frequency signal transmission, in addition to other types of devices based on magneto-mechanical coupling effects, such as energy harvesting~\cite{Halim2018} and power transferring~\cite{Du2018} systems. 

\section*{Acknowledgment}
This work was supported in part by the Defense
Advanced Research Projects Agency through the AMEBA (A MEchanically
Based Antenna) Program under Grant HR0011-17-2-0057.

\end{singlespace}

\bibliographystyle{elsarticle-num}


\end{document}